\pdfoutput=1
\documentclass[runningheads]{llncs}

\usepackage{graphicx}
\usepackage{amsfonts} 
\usepackage{amsmath}
\usepackage{amssymb}
\usepackage{siunitx}
\usepackage{soul}
\usepackage{xcolor}
\begin{document}
% NEW TITLE
\title{High fidelity fluid-structure interaction by radial basis functions mesh adaption of moving walls: a workflow applied to an aortic valve}
% ENDNEW

%\title{Advanced Radial Basis Functions mesh morphing for high fidelity Fluid-Structure Interaction with known movement of the walls: \\
%{s}imulation of an aortic valve}

%
\author{Leonardo Geronzi\inst{1} \and
Emanuele Gasparotti\inst{2}  \and
Katia Capellini\inst{2} \and
Ubaldo Cella\inst{3} \and Corrado Groth\inst{3} \and Stefano Porziani\inst{3} \and Andrea Chiappa\inst{3} \and Simona Celi\inst{2} \orcidID{0000-0002-7832-0122} \and Marco Evangelos Biancolini\inst{3} \orcidID{0000-0003-0865-5418}}

\authorrunning{Geronzi et al.}
%NEW
\titlerunning{High fidelity FSI by RBF mesh adaption} % Part of RIGHT running header
%ENDNEW
%First names are abbreviated in the running head.
%NEW
\institute{ANSYS Inc., 35-37 Rue Louis Guérin, Villeurbanne, France\\
\email{leonardo.geronzi@ansys.com} \and
BioCardioLab, Bioengineering Unit, Fondazione Toscana "G. Monasterio", Heart Hospital, Massa, Italy\\
\email{\{gasparotti,s.celi\}@ftgm.it} \and
Department of Enterprise Engineering "Mario Lucertini", University of Rome Tor Vergata, Roma, Italy\\
\email{biancolini@ing.uniroma2.it}}
%ENDNEW
\maketitle              % typeset the header of the contribution

%NEWABSTRACT
\begin{abstract}
Fluid-Structure Interaction (FSI) can be investigated by means of non-linear Finite Element Models (FEM), suitable to capture large deflections of structural parts interacting with fluids, and Computational Fluid Dynamics (CFD). High fidelity simulations are \iffalse gained thanks to \fi obtained using the fine spatial resolution of both the structural and fluid computational grids. A key enabler to have a proper exchange of information between the structural solver and the fluid one is the management of the interface at \iffalse wet \fi wetted surfaces where the grids are usually non matching. A class of applications, known also as one-way FSI problems, involves a complex movement of the walls that is known in advance as measured or as computed by FEM, and that has to be imposed at the boundaries during a transient CFD solution. Effective methods for the time marching adaption of the whole \iffalse computation \fi computational grid of the CFD model according to the evolving shape of its boundaries are required. A very well established approach consists \iffalse in \fi of a continuum update of the mesh that is regenerated by adding and \iffalse destroying  \fi removing cells \iffalse so \fi to fit the evolution of the moving walls. In this paper, starting from the work originally presented in Meshfree Methods in Computational Sciences, ICCS 2020 \cite{Geronzi}, an innovative method based on Radial Basis Functions (RBF) mesh morphing is proposed, allowing \iffalse to keep \fi the retention of the same mesh topology suitable for a continuum update of the shape. The proposed method is exact at a set of given key configurations and relies on shape blending time interpolation between key frames. The study of the complex motion of a Polymeric-Prosthetic Heart Valve (P-PHV) is presented \iffalse according to \fi using the new framework and considering as a reference the established approach based on remeshing.
%ENDABSTRACT
\end{abstract}
\section{Introduction}
In all the engineering areas of development, multi-physics analysis appears highly difficult to be carried out because of the interactions between more than one physics involved. During the achievement and the analysis of coupled systems, the way in which the shape of the object influences its performances is required to be carefully taken into account.  In the biomedical engineering field for example, the design and the evaluation of the behaviour of prosthetic valves, stent-grafts and ventricular assist devices are related to both the structural and fluid mechanics physics \cite{Marom,Roy,Avrahami}. Generally speaking, numerical meshes need to be created for the specific kind of analysis and, in a multi-physics context, two or more meshes have to be realised. In this environment, each geometric change is applied to all the numerical models involved in the analysis: such update has to be performed in a rapid way and as easy as possible. This task, usually carried out through remeshing methods, \iffalse may be also \fi may also be obtained faster with the use of mesh morphing techniques. This approach allows the changing of the shape of a meshed surface so that the topology is preserved while nodal positions are updated \cite{Biancolini,Staten}: modifications are applied on a baseline grid by moving the surface nodes and propagating displacements inside the surrounding volume mesh nodes. Concerning biomedical applications, morphing methods have been applied in both bone and cardiovascular fields. Recently, in  \cite{Biancolini_tibia}, an interactive sculpting and RBF mesh morphing approach has been proposed to address geometry modifications using a force-feedback device, while in \cite{Atasoy} RBF have been applied to improve cranioplasty applications. 
In the cardiovascular field, morphing approaches were employed in \cite{This} for the registration procedures of the cardiac muscle and to model an aorta aneurysm \iffalse carrying \fi by exploiting a one-way FSI \cite{Capellini,Capellini_2,Capellini_3}. In literature RBF have been extensively employed to tackle FSI problems \cite{Groth-2019}, using the modal method for both static \cite{Bia-2016,Bia-2020} and transient simulations \cite{Zuijlen-2007,Domenico-2018,Costa-2020}, as well as the two-way approach \cite{Cella-2012,keye-2011} with RBF-based mapping methods \cite{Biancolini-2018}. In this work the state of the art regarding biomedical one-way FSI applications, as shown in \cite{Capellini_3}, is further improved for a transient simulation by taking into account the non-linear deformations of the \iffalse wet \fi wetted surfaces during motion. An ad-hoc workflow was developed to transfer and to update, using an RBF-based morphing technique, the CFD mesh, incrementally adjusting the geometry according to the non-linear evolution predicted by the FEM solver. To demonstrate the effectiveness of the proposed approach, it was applied to a tailor-made Polymeric-Prosthetic Heart Valve (P-PHV); these devices \cite{Ghosh} proved to significantly reduce blood coagulation problems, maintaining excellent properties in term of strength, efficient function and long-term durability \cite{Bezuidenhout,Ghanbari}. This work is arranged as follows: at first, an introduction on FSI coupling is given, comparing remeshing and morphing workflows. The proposed procedure is implemented in the following paragraph, in which the incremental approach is shown in the P-PHV case. Results are finally discussed and compared \iffalse to \fi with those obtained by remeshing.

\section{FSI coupling: known-imposed motion of the walls}

One of the most sensitive processes in a Fluid-Structure Interaction analysis concerns the management of the fluid-solid interfaces; at the boundary surfaces between these two domains, solution data is shared between the fluid solver and the structural one.
In the most rigorous FSI approach, a bidirectional coupling between structural and fluid dynamic solutions is involved in an iterative process.
Such so called 2-way coupling introduces both the complexity related to the proper mapping of loads between the common boundaries of the two numerical domains (which are discretized by non-conformal meshes), and the requirement of an efficient procedure able to adapt the fluid dynamic domain to the structural solution in terms of boundaries displacement.
When the target displacement is known, or the deformation is \iffalse poorly \fi weakly affected by the aerodynamic loads, the procedure can progress in a single direction consisting    \iffalse in \fi of matching the CFD domain boundaries according to the target geometry.
This one-way coupling does not guarantee energy conservation at the fluid-solid interface but holds the benefit of lower computational time in comparison \iffalse to \fi with the bidirectional one, in which a continuous data exchange between Computational Structural Mechanics (CSM) and CFD solvers is required.
The uncertainties introduced by the mapping procedure are avoided, as well as the development of procedures governing the computations and the communications between different numerical environments.
%NEW
Several strategies are possible to link the computational domain to target shapes.
A useful technique consists in describing the geometry adopting CAD based mathematical representations of surfaces, as Non-Uniform Rational B-Splines (NURBS).
The aim is to provide a parametric formulation of the model that, adopting opportune mapping procedures, is able to reconstruct complex shapes derived from images as medical CTA data \cite{Morganti:2015}.
This approach, introduced in 2005 \cite{Hughes:2005}, is called \textit{Isogeometric Analysis} (IgA).
In \cite{Hsu:2015}, such parametric design platform is coupled with the concept of immersed boundary \cite{peskin:2002} referring to the so called \textit{Immersogeometric} FSI analysis. Immersed boundary approach, in combination \iffalse of \fi with a rigid dynamic solution of a bi-leaflets mechanical heart valve, is considered in \cite{DeTullio:2009}.
An example of a dynamic Immersogeometric analysis applied to the FSI analysis of a bio-prosthetic heart valves is reported in \cite{Xu:2018}.

The implementation of overset meshes (also called \textit{Chimera} grids) to \iffalse face \fi perform aeroelastic analyses offers several advantages.
It allows to decompose the mesh generation problem to limited portions of domain and simplifies the adoption of structured meshes which allow, with respect to unstructured/hybrid meshes, a better control of the cells clustering, the generation of high quality domains and the provision of more accurate solutions of boundary layers.
An example of the application of overset grid methodologies in fluid-structure interaction (FSI) problems is reported in \cite{Campbell:2014}. 

As introduced, a common strategy to \iffalse face \fi tackle the geometric adaptation problem consists in implementing this action directly in the numerical domain by mesh morphing techniques.
Some of the most important advantages offered by this approach are: it does not require a remeshing procedure, it allows to preserve the robustness of the computation and it can be performed within the progress of the computation.
The quality of the morphing action depends on the algorithm adopted to smooth the volume.
One of the most efficient mathematical framework to perform this task is recognized to be based on Radial Basis Functions.
%ENDNEW

In this work two Arbitrary Lagrangian Eulerian (ALE) methods \cite{Hirt,Ferziger,Zhang} for moving meshes are employed: the first based on remeshing algorithms, called in this paper \lq\lq standard\rq\rq  approach and the novel one using RBF mesh morphing procedures.

\subsection {FSI analysis based on remeshing}

\begin{figure}
\centering
\includegraphics[scale=0.25]{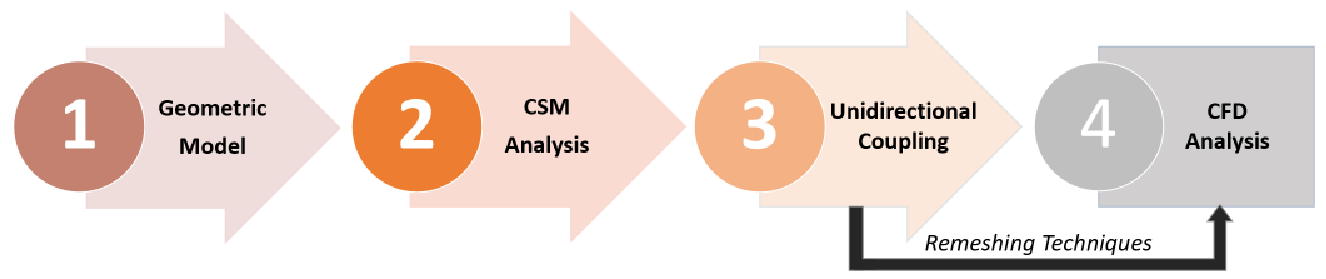}
\caption{Fluid-Structure Interaction analysis flowchart using remeshing tools.}
\label{standard}
\end{figure}

In this first kind of analysis, as shown in Fig. \ref{standard}, a specific component is responsible to transfer the deformation of the CSM mesh to the grid of the CFD solver. In fact, one of the most applied strategies to manage body meshes in FSI applications is simply to move the grid for as long as possible and, when the quality of the mesh becomes \iffalse critical \fi poor, to enable the remeshing tools updating the low-quality cells generated by high displacements in the fluid domain \cite{Baum}. \iffalse Larger are \fi The larger the displacements of the grid, \iffalse wider \fi the greater the distortions of the cells; remeshing method agglomerates cells that violate the initially defined Skewness \cite{skew} or size criteria and remeshes them. If the new cells or faces satisfy the Skewness criterion, the mesh is locally updated with the new cells, interpolating the solution from the old cells. Otherwise, the new cells are discarded and another remeshing step is required.
Obviously, when small displacements are involved and just few remeshing steps are necessary, this approach may be considered efficient. However, for larger displacements, the number of remeshing steps increases to avoid the presence of negative and invalid elements; as a result, the simulation is slowed down because a new mesh with different numbers and positions of nodes and cells has to be generated more frequently.

\subsection{A novel FSI workflow using RBF mesh morphing}

\iffalse
The high fidelity method here proposed tries to overcome the problems related to remeshing tools. It is based on the deformation of the computational grid using mesh morphing, a method to gradually change a source shape into a target one \cite{Profir}.  It aims at moving meshes with large displacements of the geometry without adopting the regeneration of the cells, typical of remeshing.
\fi

To overcome the problems related to remeshing, a procedure based on mesh morphing to adapt the shape according to a target one \cite{Profir} is \iffalse here \fi described. Among the morphing methods available in the literature, RBF are well known for their interpolation quality also on very large meshes \cite {Biancolini,Yu,Kansa}. RBF allow to interpolate everywhere in the space a scalar functions known at discrete points, called Source points (Sp). By interpolating three scalar values it is possible, solving a linear system of order equal to the number of Sp employed \cite{Biancolini}, to describe a displacement of the Sp in the three directions in space. The interpolation function is defined as follows:
\begin{equation}
\label{eq:RBF_eq_1}
s(x)=\sum_{i=1}^{N}\gamma_i \varphi \left (\left \| x-x_{s_{i}} \right \|\right) + h(x)
\end{equation}
where  \( \mathit{x} \) is a generic position in \iffalse the \fi space,  $x_{s_{i}}$ the Sp position, $s(\cdot)$ the scalar function which represents a transformation $\mathbb{R}^n$ $\rightarrow$ $\mathbb{R}$,  $\varphi$ $(\cdot)$ the radial function of order $m$, $\gamma_i$ the weight and $h(x)$ a polynomial term with degree $m-1$ \iffalse added to improve the fit assuring uniqueness of the problem and polynomial precision. \fi that can be omitted, but in this case is added in order to improve the fit, assuring uniqueness of the problem and an exact interpolation of fields of the same form of $h(x)$.

\iffalse
RBF can be used as an effective method for enabling mesh morphing on very large meshes that are typically used in advanced industrial applications \cite {Biancolini,Yu}. They consist of a very powerful tool to interpolate everywhere in the space a scalar function which defines an exact displacement of cloud of points called Source points (Sp) towards a new cloud of points named Target points (Tp). A linear system needs to be solved for sought coefficients calculation \cite{Biancolini}.  The scalar function at an arbitrary location inside or outside the domain (interpolation/extrapolation) results to be expressed by the summation of the radial contribution of each Sp and a polynomial term, added
to improve the condition number of the problem and assure polynomial precision. The interpolation function is defined as follows:
\begin{equation}
\label{eq:RBF_eq_1}
s(x)=\sum_{i=1}^{N}\gamma_i \varphi \left (\left \| x-x_{s_{i}} \right \|\right) + h(x)
\end{equation}
where x is a generic position in the space,  $x_{s_{i}}$ the Sp position, $s(\cdot)$ the scalar function which represents a transformation $\mathbb{R}^n$ $\rightarrow$ $\mathbb{R}$,  $\varphi$ $(\cdot)$ the radial function, $\gamma_i$ the weight and $h(x)$ a polynomial.
\fi

The unknowns of the system, namely the polynomial coefficients and the weights $\gamma_i$ of the radial functions, are retrieved by imposing the passage of the function on the given values and an orthogonality condition on the polinomials. If the RBF is conditionally positive
definite, it can be demonstrated that a unique interpolant exists and in 3D, if the order is equal or less than 2, a linear polynomial in the form $h(x)=\beta_1+\beta_2x+\beta_3y+\beta_4z$ can be used. The linear problem can be also written in matrix form:

\begin{equation}
\label{rbf_matrix_form}
\begin{bmatrix}
\textbf{M} & \textbf{P}\\
\textbf{P}^T & \textbf{0}
\end{bmatrix}
\left\lbrace
\begin{matrix}
\boldsymbol{\gamma}\\
\boldsymbol{\beta}
\end{matrix}
\right\rbrace
=
\left\lbrace
\begin{matrix}
\textbf{g}\\
\textbf{0}
\end{matrix}
\right\rbrace
\end{equation}
in which $\textbf{M}$ is the interpolation matrix containing all the distances between RBF centres $\textbf{M}_{ij} = \varphi \left (\left \| x_i-x_j \right \|\right)$, $\textbf{P}$ the matrix containing the polynomial terms that has for each row $j$ the form $\textbf{P}_j = [
\begin{matrix} 1 & x_{1j} & x_{2j} & ... & x_{nj} \end{matrix} ] $ and $\textbf{g}$ the known values at Sp. The new nodal positions, if interpolating the displacements, can be retrieved for each node as:
\begin{equation}
\label{eq:RBF_eq_4}
\boldsymbol{x_{{node}_{new}}}=\boldsymbol{x_{node}} + \begin{bmatrix}
s_x(\boldsymbol{x_{node}})\\ 
s_y(\boldsymbol{x_{node}})\\ 
s_z(\boldsymbol{x_{node}})
\end{bmatrix}
\end{equation}
where $\boldsymbol{x_{{node}_{new}}}$ and $\boldsymbol{x_{node}}$ are the vectors containing respectively the updated and original positions of the given node.
\iffalse
If a deformation vector field has to be fitted in 3D (space morphing), considering $h(x)$ as a linear polynomial made up of known $\beta$ coefficients:

\begin{equation}
\label{eq:RBF_eq_2}
h(x)=\beta_1+\beta_2x+\beta_3y+\beta_4z
\end{equation}

each component of the displacement prescribed at the Source points can be interpolated as follows:
\begin{equation}
\label{eq:RBF_eq_3}
\left\{\begin{matrix}
s_x(x)=\sum_{i=1}^{N}\gamma_i^x\varphi\left (\left \| x-x_{s_i} \right \|\right)+ \beta_1^x + \beta_2^xx+\beta_3^xy+\beta_4^xz\\ 
s_y(x)=\sum_{i=1}^{N}\gamma_i^y\varphi\left (\left \| x-x_{s_i} \right \|\right)+ \beta_1^y + \beta_2^yx+\beta_3^yy+\beta_4^yz\\ 
s_z(x)=\sum_{i=1}^{N}\gamma_i^z\varphi\left (\left \| x-x_{s_i} \right \|\right)+ \beta_1^z + \beta_2^zx+\beta_3^zy+\beta_4^zz
\end{matrix}\right.
\end{equation}
The field of system (\ref{eq:RBF_eq_3}) is applied to process all the nodal positions to be updated according to the following condition:
\begin{equation}
\label{eq:RBF_eq_4}
x_{node_{new}}=x_{node} \begin{bmatrix}
s_x(x_{nodes})\\ 
s_y(x_{nodes})\\ 
s_z(x_{nodes})
\end{bmatrix}
\end{equation}

each component of the displacement prescribed at the Source points can be interpolated by means of a linear system, that can be applied to process all the nodal positions to be updated according to the following condition:
\begin{equation}
\label{eq:RBF_eq_4}
\textbf{x_{node_{new}}}=\textbf{x_{node}} \begin{bmatrix}
s_x(\textbf{x_{nodes}})\\ 
s_y(\textbf{x_{nodes}})\\ 
s_z(\textbf{x_{nodes})}
\end{bmatrix}
\end{equation}

\fi

\iffalse Being the problem solved \fi As the problem is solved pointwise, the approach is meshless and able to manage every kind element (tetrahedral, hexahedral, polyhedral and others), both for surface and volume mesh smoothing ensuring the preservation of their topology.

\iffalse
Previous equations show the meshless nature of the approach because the deformed position only depends on the original position to be moved; this means that RBF are able to manage every kind of mesh element type (tetrahedral, hexahedral and polyhedral) and the RBF fit can be efficiently used for surface and volume mesh smoothing, ensuring the preservation of their topology, namely the same number of nodes and cells with the same typology.
\fi

\begin{figure}
\centering
\includegraphics[scale=0.28]{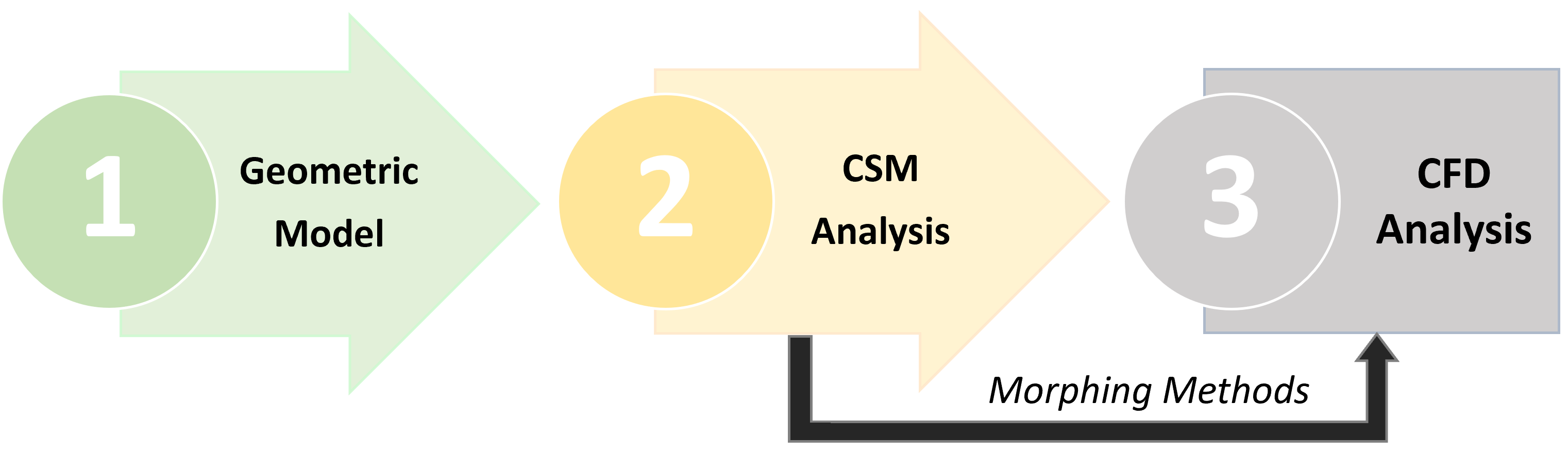}
\caption{Novel FSI workflow based on morphing techniques.} \label{New}
\end{figure}

In general, a morphing operation can introduce a reduction of the mesh quality but a good morpher has to minimize this effect and to maximize the possible shape modifications. If mesh quality is well preserved, morphing has a clear benefit over remeshing because it avoids introducing noise \cite{Biancolini}. This procedure provides for the sampling of the displacements of the structural model. In this way, the surface nodes of the FE model are used as Sp to apply RBF mesh morphing in the space.
The detailed workflow shown in Fig. \ref{New}, in addition to the realization of the geometric model, provides for three main steps:

\begin{enumerate}
\item[-] CSM settings
\item[-] RBF Morph application
\item[-] CFD set-up
\end{enumerate}

\subsubsection{CSM settings.}

Given that in this kind of FSI analysis the deformations of the fluid grid are imposed by the motion of the walls of the CSM domain, a structural simulation has to be conducted to obtain the deformed positions in the space during time. During the structural simulation, Source points ($S_p$) and Target points ($T_p$), that are the surface nodal positions of the grid in this case, are recursively extracted and stored in files. These files are structured as a list of nodes and the related displacements to obtain the nodes in the following sampled configuration. 

\subsubsection{RBF Morph application.}

To apply the shape deformation without remeshing, the morpher RBF Morph is in this study employed \cite{Biancolini_2}. To delimit the morphing action, an encapsulation technique has to be implemented, defining in this way sub-domains or parts of the fluid domain within which the morpher action is applied \cite{RBF_user}.  According to \cite{rbf_guide}, $S_p$ effectively used are a random subgroup which includes a percentage close to 2\% of the total extracted surface nodes of the structural model. A Radial Basis Function is selected and, when the settings of the procedure are completed, morphing solutions are calculated, each per morphing transitions, from the starting configuration to the final position of the model. Every transition, a new deformed mesh is generated and the solution files are stored.

\subsubsection{CFD set-up.}

Once all the morphed meshes are obtained, the CFD analysis is carried out every time step on a new deformed mesh, by using a specific script containing the RBF functions to update the nodal positions. \iffalse in Scheme Programming Language. This script is recalled \fi The script is run before the starting of the CFD simulation and it allows the automatization of the MultiSol tool of RBF Morph software in order to guarantee the synchronous implementation of the deformed shape in fluid environment.
The nodal displacement inside two subsequent morphed positions is simply handled inside the Scheme script by means of a parameter called amplification A(t) which varies over time from 0 to $A_0$ (usually equal to +1 or -1).
\iffalse Because of the samplings \fi Due to the sampling data available at different times, an accurate and effective solution can be obviously achieved, handling all the morphing deformations separately and linearly modulating the amplification within each of the sampled grids: calling $i$ the index that concerns all the mesh conformations $P_0$...$P_{n}$, the amplification $A_i(t)$ inside each time step can be evaluated as:

\begin{equation}
\medskip
\label{eq:lin_amplification}
\centering
A_{i}(t) = \begin{cases} 0, & \mbox{if } t\leq t_{i}\\ \frac{t-t_{i}}{t_{i+1}-t_{i}}, & \mbox{if } t_i<t<t_{i+1}\\ 1, & \mbox{if } t\geq t_{i+1}\end{cases}
\medskip
\end{equation}

In this way, the fluid grid is able to follow the displacements of the CSM mesh and solution can be quickly calculated every time step on a new deformed grid. 
%NEW
In other situations, not only linear amplifications but also quadratic and sinusuoidal ones can be used to manage motion through morphing, to replicate more correctly the studied dynamics.

\section{Implementation of the workflows to study the opening phase of a polymeric aortic valve}

In this paper, both the presented FSI solution approaches are applied to study the opening phase of a P-PHV. This kind of analysis is conducted on a prosthetic valve to study its haemodynamic behaviour during the opening phase in which the leaflets or cusps \cite{Anderson,Iaizzo} are pushed open to allow the ejection of the blood flow and to investigate the effects of the solid domain (P-PHV) \iffalse towards \fi on the fluid domain (blood flow).

\begin{figure}[ht]
\centering
\includegraphics[scale=0.625]{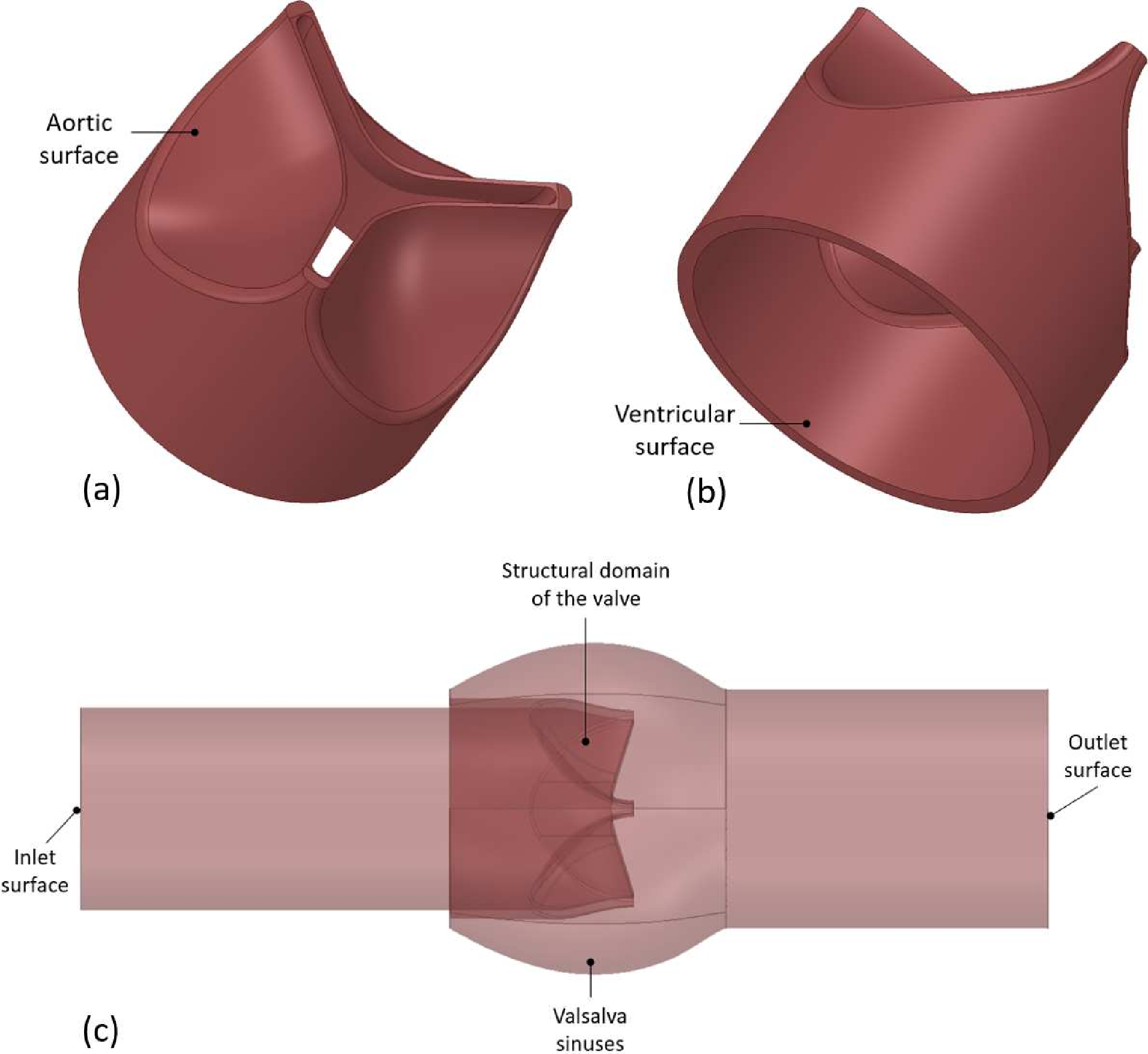}
\caption{CAD model of the Polymeric-Prosthetic Heart Valve. Indications of aortic side (a) and ventricular side (b). Blood domain with the hollow volume of the valve (c).} \label{CAD_T}
\end{figure}

The design of a prosthetic aortic valve is realised using the software SpaceClaim, following the geometry proposed in \cite{Jiang}, and then imported inside ANSYS Workbench (v193).
Fig. \ref{CAD_T} depicts the CAD model of the polymeric valve here investigated and  the fluid domain of the blood in which also the Valsalva sinuses are represented \cite{sinuses}. This fluid domain is obtained by a boolean intersection between the solid domain of the valve and a vessel made up of an inlet and outlet tube.

The computational grid of this structural model is made up of 300,000 tetrahedral elements and it is obtained inside ANSYS Meshing. The material chosen to model the valve is isotropic linear elastic with a Young's modulus of 3 MPa, and a Poisson's ratio of 0.4. The mesh of the fluid domain is obtained inside Ansa (v20.0.1) and made up of \iffalse 1564792 \fi 1.56 million tetra-hexaedral elements with a maximum starting Skewness (Sk) of 0.694. A velocity inlet and a systolic aortic pressure outlet varying over time boundary conditions have been provided. The flow model selected is Viscous-Laminar and the fluid is considered as incompressible and Newtonian with a viscosity of 4 cP.

\subsection{Study of the valve using the remeshing workflow}

\begin{figure}[!ht]
\centering
\includegraphics[scale=1.11]{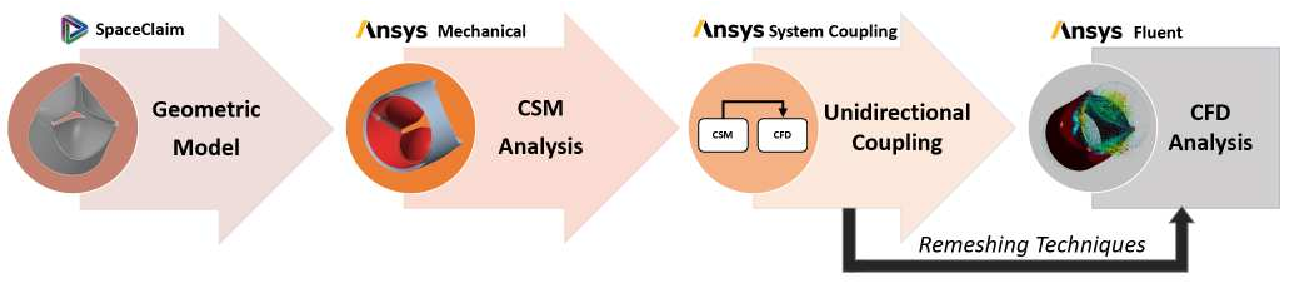}
\caption{Flowchart of the FSI analysis implemented inside ANSYS Workbench to study the behaviour of a P-PHV using Remeshing Method.} \label{standard_valve}
\end{figure}

The workflow of the first approach based on remeshing and implemented inside ANSYS Workbench is proposed in Fig. \ref{standard_valve}. In this case, the displacements of a transient CSM in ANSYS Mechanical (v193) are transferred to the fluid domain inside ANSYS Fluent (v193) by means of System Coupling.  The body of the prosthetic device is fixed \iffalse in \fi at its bottom surface. \iffalse, i.e. the circular ring in the ground.\fi  Concerning the loads, a physiological time-varying pressure is uniformly applied to the ventricular portion of the valve to simulate the opening pressure.

Fluent was used as the CFD solver, \iffalse activating the Dynamic Mesh \fi the Dynamic Mesh option was activated to deform and regenerate the computational grid \cite{Ansys_guide}: the \iffalse Spring-Laplace \fi Spring-based Smoothing Method and the Remeshing Method, two tools to solve the analysis of deformed domains due to boundary movement over time \cite{Si}, are applied.

%NEW
Inside the System Coupling component, all the surfaces of the valve in contact with the blood are set as fluid-solid interfaces and a time step of \iffalse 1e-5 \fi \num{e-5} s is selected. For the run, 48 cores of \textit{Intel(R) Xeon(4) Gold 6152 @2.10 GHz} and a 256 Gb RAM are used.
%ENDNEW

\subsection{Study of the valve using the morphing approach}
\begin{figure}
\centering
\includegraphics[scale=0.31]{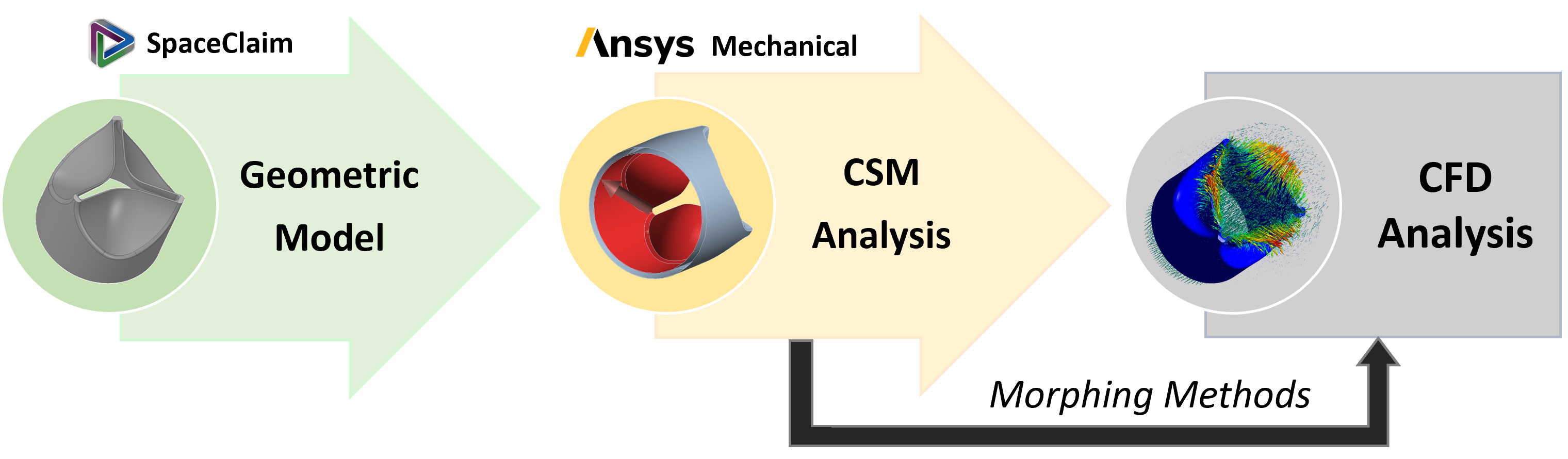}
\caption{Flowchart with the application of morphing.} \label{New_valve}
\end{figure}

The novel workflow, based on a RBF mesh morphing, also in this case developed into the ANSYS Workbench environment is shown in Fig. \ref{New_valve} which depicts the overall  incremental procedure; in particular, the System Coupling component, fundamental in the previously shown approach, \iffalse here \fi is not employed here.

\begin{figure}[ht]
\centering
\includegraphics[scale=1.83]{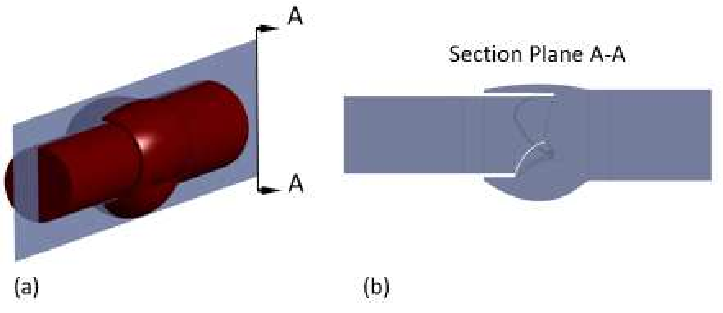}
\caption{Fluid domain with the section plane A-A (a); section plane A-A (b)} \label{section-p}
\end{figure}

%NEW
Results obtained with the use of RBF Morph software are compared \iffalse to \fi with those achieved using Fluent remeshing tools. They are reported in terms of pressure map, velocity streamlines, 2D velocity contours on a section plane A-A (Fig. \ref{section-p})  and computational time.

%ENDNEW

\subsubsection{Structural simulations.} %NEW
The same CSM analysis implemented in the remeshing approach with the same time-varying physiological pressure is conducted. A specific Application Customization Toolkit (ACT) Extension, i.e. a method to achieve custom applications inside ANSYS Mechanical, is developed to extract the deformed configurations of the leaflets and collect Sp and Tp according to a sampling time chosen by the user: in this %NEW
first case, ten different surface nodal positions  ($P_0$...$P_9$) are collected sampling every 2 ms (0-18 ms).

\subsubsection{RBF Morph set-up.}

\begin{figure}[ht]
\centering
\includegraphics[scale=0.45]{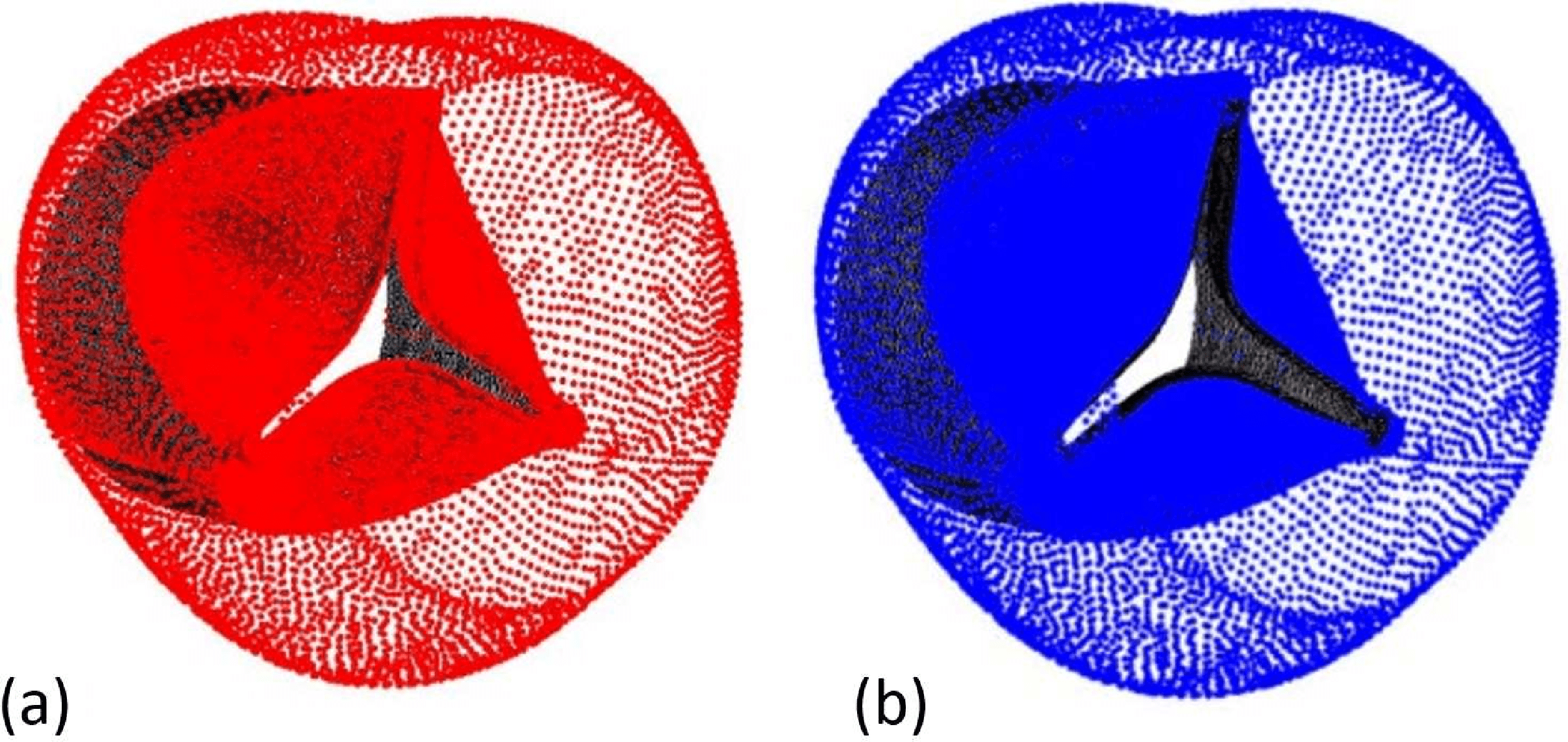}
\caption{First morphing step of the opening: Sp on the surface of the valve and fixed Surf Set on the wall of the Valsalva sinuses (a), Tp on the surface of the prosthetic device to obtain a first movement keeping stationary the points on the STJ surface (b).} \label{SP-TP}
\end{figure}

The RBF Morph Fluent (v193) Add-On is adopted: the RBF are used to step by step project  the $S_p$ to the $T_p$ and interpolate the displacements of the volume nodes inside the blood domain over time. RBF Morph tools may impose a blood domain adaptation in the CFD analysis following the deformed shapes of the valvular domain, extracted from the previous CSM simulation. In the case of the valve, to control the morphing action in the space, no encapsulation volumes \cite{RBF_user} are generated but a fixed Surf Set (with  null displacement during the movement of the leaflets)   on the wall of the Sinotubular Junction (STJ) \cite{Maselli} is preferred (Fig. \ref{SP-TP}). In this way, since the distortion of the mesh is extremely high, such deformations can be distributed also in the other zones of the fluid domain and not only in proximity to the valve. 

\begin{figure}[!t]
\centering
\includegraphics[scale=0.5]{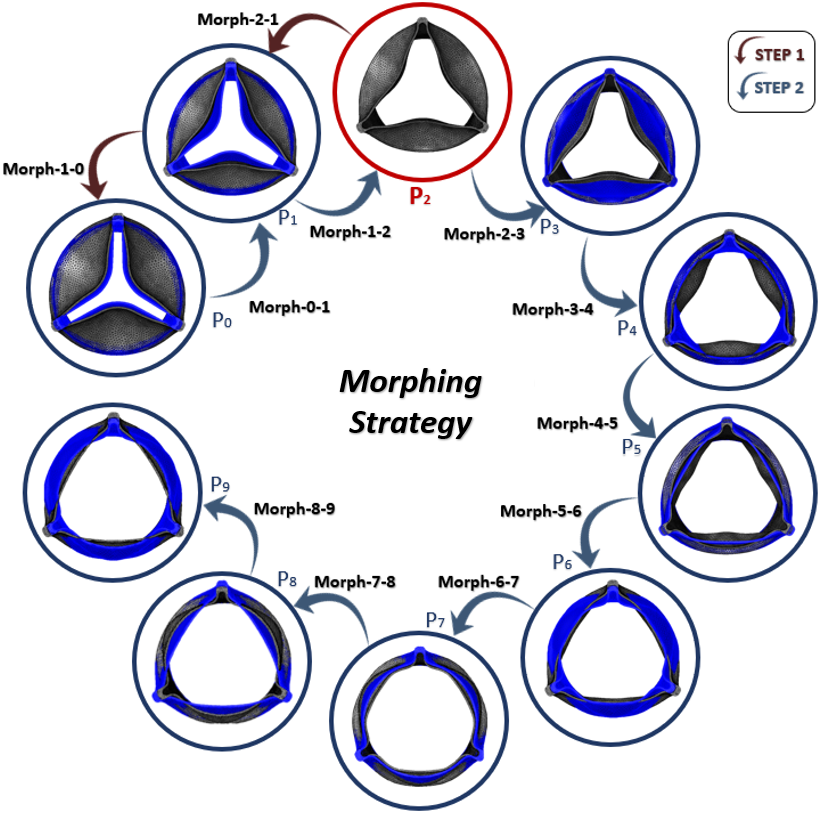}
\caption{Backwards/Forwards strategy: step 1 represents a backwards morphing procedure from $P_2$ (partially open position) to $P_0$ (initial position)  while step 2 allows to reach position $P_9$ (completely open) starting from position $P_0$. This strategy \iffalse allows to prevent \fi prevents excessive mesh quality degradation.} \label{morphate}
\end{figure}

To prevent mesh quality reduction, in this work the bi-harmonic kernel $\varphi(r)=r$ is used: it is the RBF that guarantees to keep the mesh quality degradation at minimum \cite{Bi-h}.
The nine solutions are calculated using the Solve Panel of the software, each per morphing action. 
Two different strategies are evaluated: 
\begin{itemize}
    \item \textit{Forwards}: from $P_0$ (initial position) directly to $P_9$ (totally open position). 
    \item \textit{Backwards/Forwards} (Fig. \ref{morphate}): extracting a new already deformed valve in position $P_2$ of which a new mesh is achieved, from $P_2$ going back to $P_0$ (backwards - step 1) and then from $P_0$ moving on to $P_9$ (forwards - step 2).
\end{itemize}

The first strategy produced an excessive mesh distortion with negative cells volume in particular at the centre of the leaflets. No mesh problems are detected with the adopted second approach %NEW
as it started from a robust mesh in intermediate configuration that was modified partly to return to the initial position and partly to reach the full opening phase. 
\subsubsection{Fluent setting.}

The same time step of the standard approach (\num{e-5} s) is selected. The flow is initialized with the valve in position $P_0$ and the Scheme file with the amplification of each of the nine different transitions is \iffalse recalled \fi inside Fluent.
%NEW
The amplification used for the displacements between $P_1$ and $P_9$ is linear as shown in (\ref{eq:lin_amplification}). Instead, the use of a quadratic amplification $A_0(t)$ between $P_0$ and $P_1$ is preferred in order to more realistically replicate the rapid opening of the prosthesis:
\begin{equation}
\medskip
\label{eq:quad_amplification}
\centering
A_{0}(t) = \begin{cases} 0, & \mbox{if } t = 0\\ (\frac{t}{t_{1}})^2, & \mbox{if } 0<t<t_{1}\\ 1, & \mbox{if } t\geq t_{1}\end{cases}
\medskip
\end{equation}
%ENDNEW
Using \iffalse Thanks \fi to this workflow, the FSI simulation can be conducted only by means of this CFD analysis in which the mesh is updated every time step.

%NEW SUBSECTION
\subsection{Full systolic movement using RBF mesh morphing}
\begin{figure}[!ht]
\centering
\includegraphics[scale=0.85]{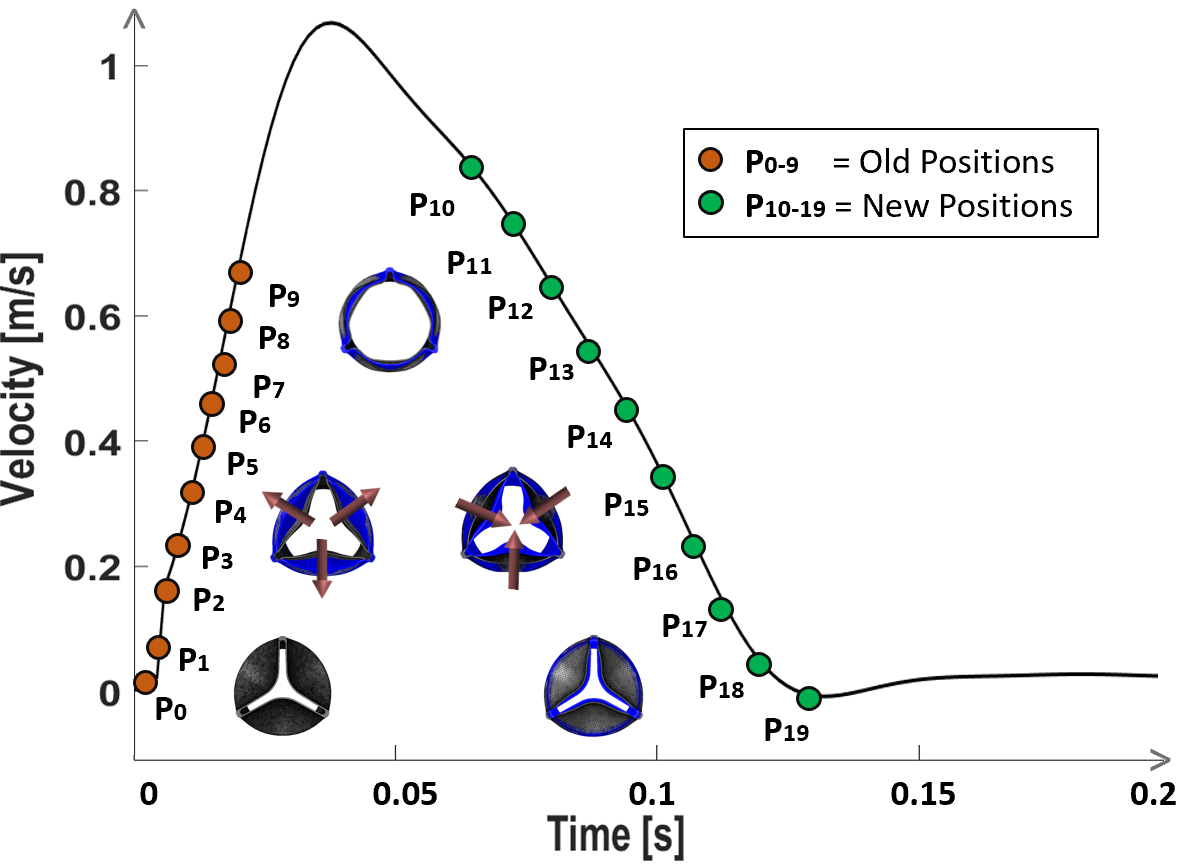}
\caption{Valvular positions sampled during the systolic phase referred to the velocity of blood flow for the inlet.} \label{total_morphing}
\end{figure}

\begin{figure}[!ht]
\centering
\includegraphics[scale=1.03]{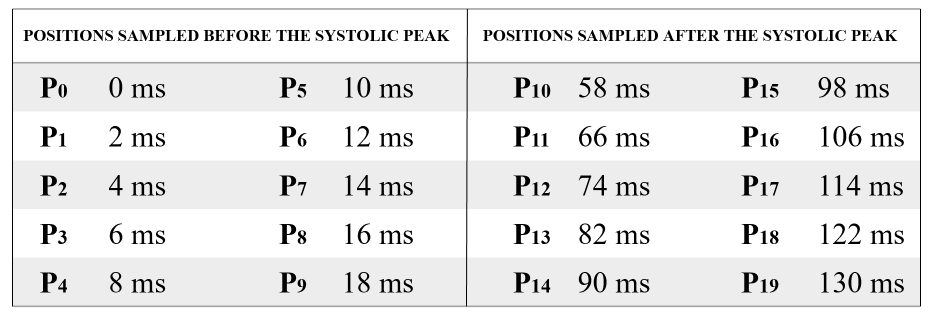}
\caption{Set of all the samples collected over time. } \label{table_pos}
\end{figure}

Given that morphing is found to be several times faster than remeshing tools, a complete opening cycle of the valve for a total of 130 ms is simulated, allowing the valve not only to open but also to provide for all the cardiac ejection and return to a position close to the initial one. For these reasons, a new structural simulation is carried out on the same 130 ms to allow the extraction of further surface nodes during time: ten additional valvular positions are sampled after the systolic peak, corresponding to the descending portion of the inlet velocity curve reported in Fig. \ref{total_morphing}, in which the valve begins closing and it comes back in a position similar to the initial one. In Fig. \ref{table_pos}, the time in which the morphing positions are sampled is reported in detail. 
Once opened, the P-PHV starts weakly swinging; all these small vibrations can be neglected in term of variation of Geometric Orifice Area (GOA), as shown in \cite{Pfensig,Maleki}. This means that the conformation of the valve between $P_{9}$ and $P_{10}$ remains quite similar, that's why no samples are taken in this interval (18 ms - 58 ms).
Moreover, since the movement of the valve is slower in the section between the systolic peak and the beginning of the diastole than the more abrupt opening at the beginning of the systole, as reported in \cite{Zakerzadeh}, positions samplings may take place every 8 milliseconds and not every 2, thus saving computational time for morphing transformations. 

Time step selected for this simulation is \iffalse 5e-4 \fi \num{5e-4} s.

%ENDNEW

\section{Results and Discussions}
\subsection{Remeshing and morphing results}
The first important parameter to check in the implementation of both the approaches is the Skewness: no negative cells are detected and maximum Sk values are 0.858 and 0.961 respectively for the procedure implemented through remeshing and morphing. No convergence issues due to mesh degradation are recorded.

%NEW
In this work, pressure and velocity results of the 3D-model at a specific and significant time step are shown using the CFD-Post software. In addition, a case comparison between the velocity contours on the A-A section plane at 4 different time instants is carried out to evaluate the differences between the two techniques presented.
%ENDNEW
\begin{figure}[!ht]
\centering
\includegraphics[scale=0.98]{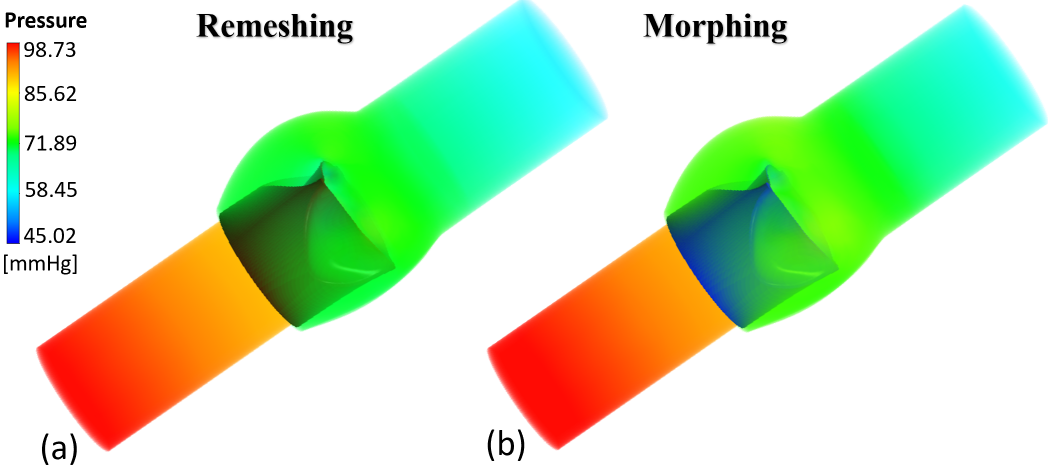}
\caption{Volume rendering of the pressure at t=7 ms, when the valve is for the first time open: remeshing approach (a), morphing approach (b).} \label{pressure}
\end{figure}

\textit{Pressure map -}   A 3D-pressure rendering at the time in which the valve is for the first time open (7 ms)  is represented in Fig. \ref{pressure}; as \iffalse it \fi can be observed, no significant differences in terms of pressure values are recorded between the two approaches: the mean pressure are 69.64 mmHg and 70.31 mmHg respectively for the simulation based on remeshing and the analysis using mesh morphing (difference of less than 1\%).

\begin{figure}[ht]
\centering
\includegraphics[scale=0.14]{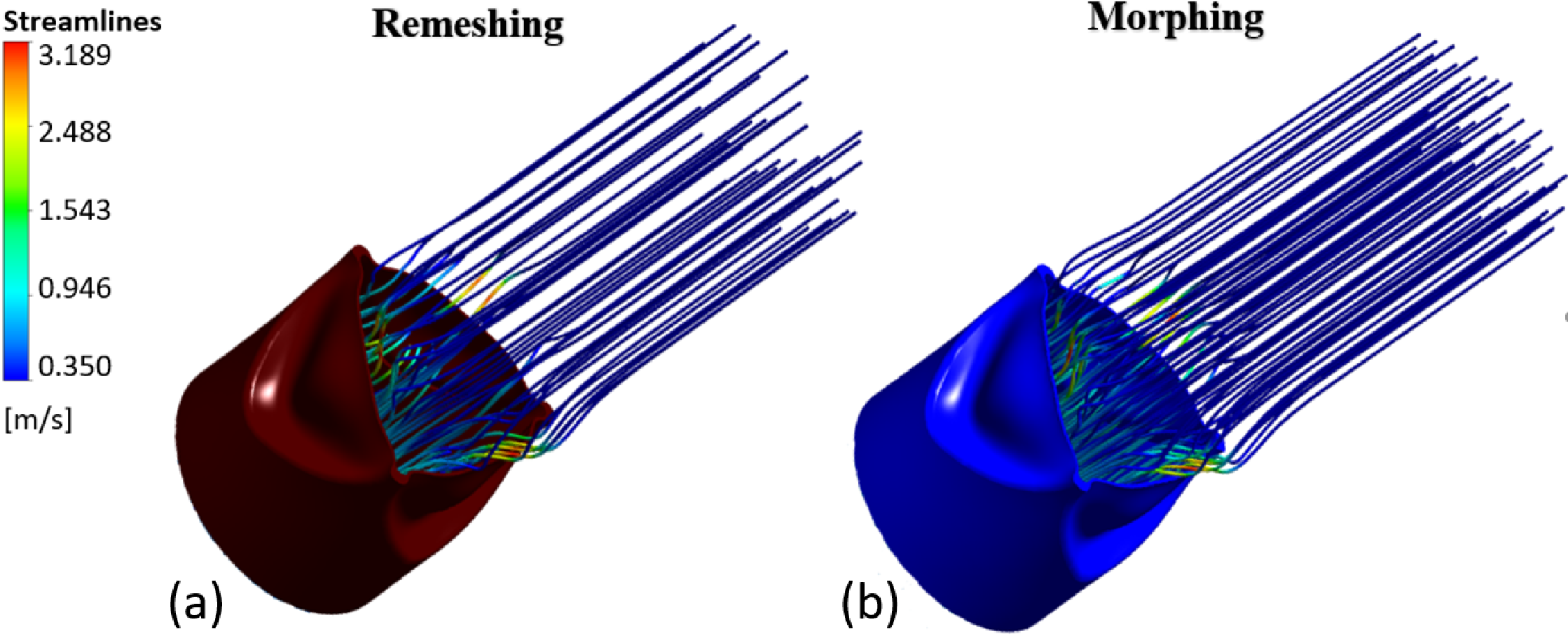}
\caption{Streamlines at t=7 ms:  FSI using remeshing (a),  FSI using morphing (b).} \label{Streamlines}
\end{figure}

\textit{Velocity streamlines -} Streamlines of both the solution methods are reported in Fig. \ref{Streamlines}. It is possible to observe how in this 3D-representation the local maximum values are placed in the same zones. Maximum local fluid velocities are reached right at t=7 ms because the fast movement of the leaflets pushes away the blood: maximum value for the FSI implemented with the novel approach is 3.011 m/s, about 5.6\% less than the value detected in the analysis based on remeshing, that is 3.189 m/s. 

\begin{figure}[h!]
\centering
\includegraphics[scale=1.2]{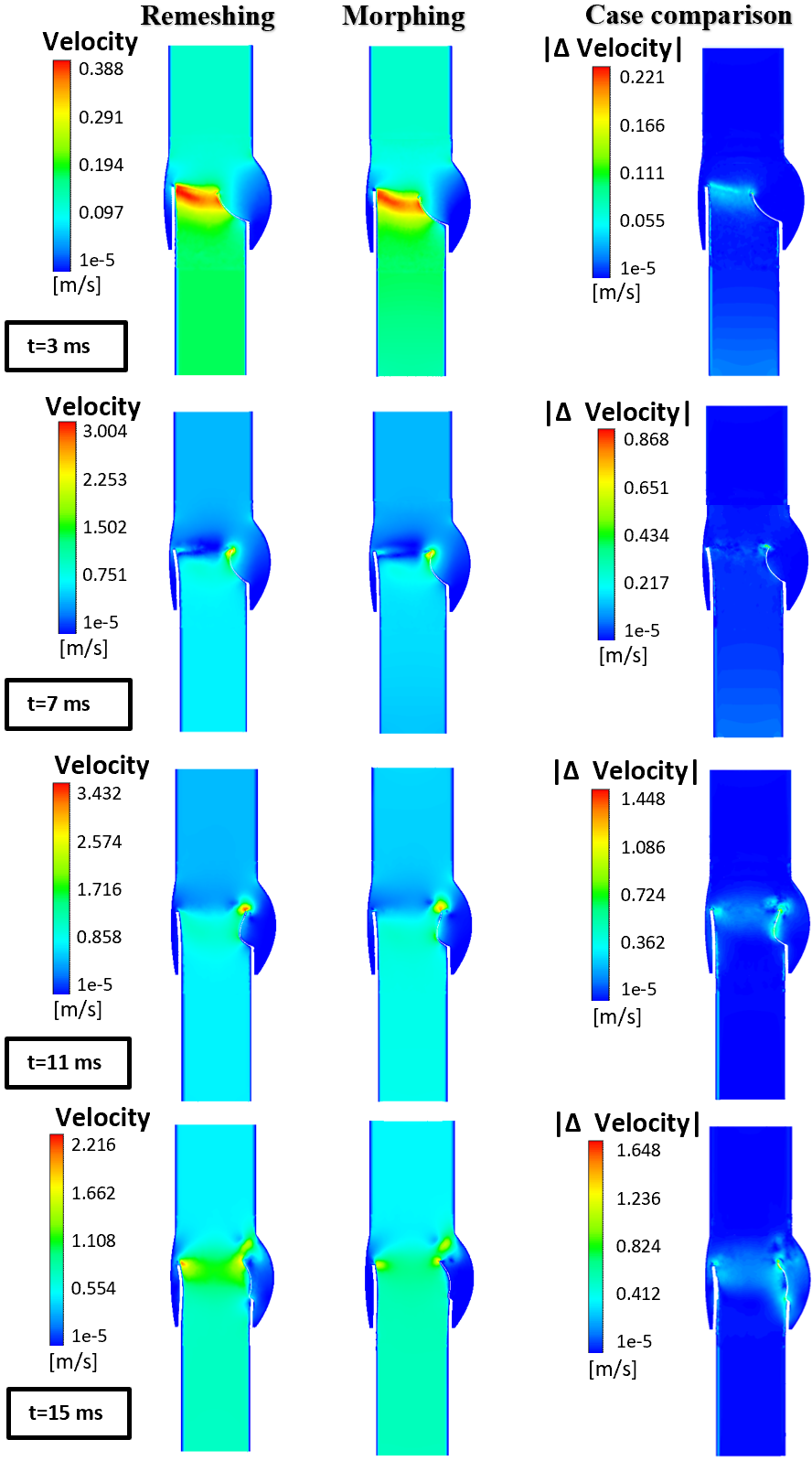}
\caption{Velocity contours during the opening of the valve: results of 4 different time instants  for remeshing, morphing and case comparison obtained as absolute value of the subtraction.} \label{contour-p}
\end{figure}

%NEW
\textit{Velocity contours -} To investigate in more depth the differences between remeshing and morphing, velocity contours on the plane passing through the centre of the valve, shown in Fig. \ref{section-p}, are evaluated. Four different evenly spaced time steps are here reported (Fig. \ref{contour-p}): 3 ms, 7 ms, 11 ms, 15 ms. To better identify these differences, in the same picture the case comparison by exploiting the subtraction in absolute value between the velocity of remeshing and morphing is also evaluated. The numerical results of interest are reported in Table \ref{tab:table_res}. Here, maximum and average velocity values on the section plane A-A for both the approaches are evaluated. The local maxima in both cases are significantly above the physiological values normally found. This is due to the motion imposed on the valve which pushes strongly fluid close to the leaflets during opening. Moreover, the high difference that emerges from the image with the case comparison is mainly attributable to the difference between local maxima moved slightly in the 2-D space. The maximum difference detected between the maximum values occurs after 11 milliseconds when the morphing value is 6.4$\%$ lower than that of remeshing. In the case of the average value, the highest gap is at t=15 ms: in this case, the FSI carried out by means of morphing returns a value 0.063 times lower than the simulation with remeshing. However, despite a very high stretching of the elements, there are no marked irregularities in the case of mesh morphing.

\begin{table}[h!]
\begin{center}
\caption{Values obtained for the analysis of velocity on the section plane A-A.}
\begin{tabular}{|c|c|c|c|}
\hline
TIME & Max. velocity remeshing & Max. velocity morphing & Difference between Max.\\
\hline
t=3 ms & 0.388 & 0.381 & -0.007 (-1.8$\%$) \\
%\hline
t=7 ms &  3.004 & 2.937 & -0.067 (-2.2$\%$)\\
%\hline
t=11 ms  & 3.432 & 3.211 & -0.221 (-6.4$\%$)\\
%\hline
t=15 ms & 2.216 & 2.088 & -0.128 (-5.8$\%$)\\
\hline
\end{tabular}

\vspace{0.5cm}

\begin{tabular}{|c|c|c|c|}
\hline
TIME & Ave. velocity remeshing & Ave. velocity morphing & Difference between Ave.\\
\hline
t=3 ms & 0.081 & 0.079 & -0.002 (-2.4$\%$)\\
%\hline
t=7 ms &  0.289 & 0.294 & +0.005 (+1.7$\%$)\\
%\hline
t=11 ms  & 0.456 & 0.434 & -0.022 (-4.8$\%$) \\
%\hline
t=15 ms & 0.788 & 0.738 & -0.050 (-6.3$\%$)\\
\hline
\end{tabular}
\label{tab:table_res}
\end{center}
\end{table}

%ENDNEW

\textit{Computational time -} The time required to run the simulation \iffalse calculate the solution \fi is the most interesting difference between these two solution methods: with the same number of elements and \iffalse to \fi parity of time step (\num{e-5} s), the standard workflow requires 6283 minutes to solve the problem while the CFD analysis with the morphing application only 396 minutes, approximately 16 times faster.
Furthermore, if time step is increased, Fluent solver using remeshing starts having troubles and the obvious consequence is the non convergence of the model. This phenomenon does not happen with morphing \iffalse by \fi for which the time step can be \iffalse highly \fi increased significantly without problems and solution time can be reduced up to 60 times compared \iffalse to \fi with that of remeshing.

%NEW SUBSECTION
\subsection{Global systolic phase simulated through mesh morphing}
\begin{figure}[ht]
\centering
\includegraphics[scale=1.0]{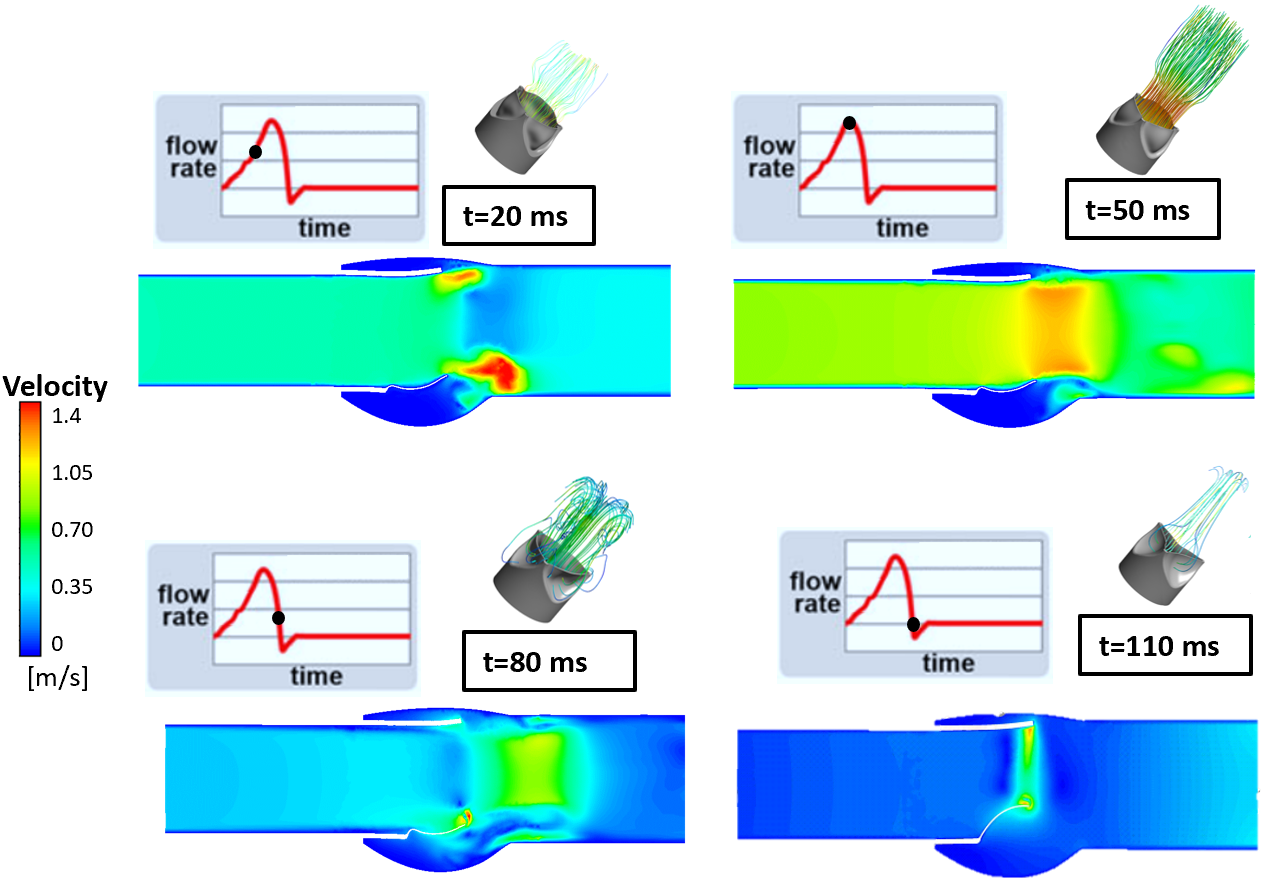}
\caption{Four velocity contours during all the systole and the corresponding valvular positions with the flow lines.} \label{tot_morphing}
\end{figure}

The simulation with the motion of the walls imposed on the entire systolic phase is correctly completed by means of mesh morphing. The velocity contours computed on plane A-A (Fig. \ref{section-p}) and the 3D representations of the P-PHV are shown in Fig. \ref{tot_morphing}. As it can be seen, after 20 ms, the propagation of the blood flow is still strongly affected by the abrupt opening of the valve itself and the local maxima of velocity are located exclusively in proximity of the leaflet. Near the systolic peak instead, the highest velocity region is more evenly distributed in the central orifice area of the valve as the leaflets have now completed their opening and no longer push fluid against the sinuses of Valsalva. At t=80 ms, passed the flow peak, the valve is already starting to close reducing its orifice area while after 110 ms the prosthesis is very close to achieving the final configuration: in this case, as it can be seen, the 2D velocity contour distribution is very similar to the one shown after 3 ms in Fig. \ref{contour-p}.

By means of the new time step and through mesh morphing, 311 minutes are required to run the simulation.

%ENDNEW

\section{Conclusions}
In this work, a fast high fidelity workflow for a multi-physics Fluid-Structure Interaction analysis exploiting ANSYS Workbench and RBF Morph is presented.  In particular, this powerful system has proved to be useful and accurate in all those applications \iffalse in \fi for which the movement of the fluid domain is strongly controlled by the deformation of the solid one
%NEW
as in the case of valves or bodies, in contact with a fluid, whose motion is known or imposed.
%ENDNEW
The possibility to replicate the non-linear analysis in which the motion of a solid model is transferred to a fluid domain carefully handling the interfaces is demonstrated. In a specific application case, the attempt to transfer aortic valve opening kinematics to the CFD analysis through a mesh morphing technique \iffalse results to be \fi is successful and the effectiveness of the FSI simulation implemented through mesh morphing  to reproduce the fluid dynamic field of the aortic valve is here shown and validated.  Morphing  has turned out to be very consistent in comparison \iffalse to  \fi with remeshing. Further research is needed to asses in more detail the differences between these two distinct approaches. 
%NEW 
It is clear that in order to reproduce the dynamics of the blood flow even more faithfully, a 2-way FSI should be set up. However, further studies are in progress to propose this 1-way approach to realise a 2-way Fluid-Structure Interaction where the exchange of information between the two different physics occurs at the end of each cardiac cycle and not time step by time step. In this way, the cost of the 1-way solution would be preserved and repeated over several cycles with the appropriate update of the boundary conditions.
%ENDNEW
The key point is that this novel workflow on a 1-FSI analysis about the study of the opening of a polymeric aortic valve has shown a reduction of the simulation time up to 16 times in comparison \iffalse to \fi with that required by remeshing methods using the same time step.
%NEW
Furthermore, this important saving in computational time and the possibility of increasing the time step have made it possible to study the whole systole, managing the full movement of the valve during this phase.
%ENDNEW
Until now, numerical FSI simulations in medical field have been used in a restricted manner precisely because of the long time required to solve them, as in the case of heart valve computational analyses. Dealing with clinical trials, this approach could ensure a considerable time saving with a decrease of the final mesh quality which results acceptable for the numerical solver here adopted. This new workflow has definitely broken down this limit, granting in this case the possibility to test several models of P-PHV in a lot of flow conditions. In addition, a parametric design of P-PHV combined with this method could finally make a strong contribution to the patient-specific aortic valve replacement.

\section{Acknowledgment}
The research has received funding from the European Union’s Horizon 2020 research and innovation programme under the Marie Skłodowska-Curie grant agreement No 859836, MeDiTATe: ``The Medical Digital Twin for Aneurysm Prevention and Treatment", and has been partially supported by RBF Morph\textsuperscript\textregistered.



\end{document}